\documentclass[review,floatfix]{elsarticle}

\usepackage{lineno,hyperref}
\modulolinenumbers[5]

\def\D{{\rm d}}
\def\E{{\rm e}}
\def\I{{\rm i}}
\usepackage{mathtools}
\usepackage{amsmath}
\usepackage{amssymb}
\usepackage{color}
\usepackage{graphicx}
\usepackage{dcolumn}
\usepackage{times}
\usepackage{soul}
\usepackage{afterpage}
\usepackage{mathabx}

\journal{~}

\begin{document}

\begin{frontmatter}

\title{Reshaping of a Janus ring}

\author{A.~P. Zakharov, L.~M. Pismen\fnref{e-mail}}
\address{Department of Chemical Engineering, 
Technion -- Israel Institute of Technology,\\ Haifa 32000, Israel}
\fntext[e-mail]{pismen@technion.ac.il}




\begin{abstract}

We consider reshaping of closed Janus filaments acquiring intrinsic curvature upon actuation of an active component -- a nematic elastomer elongating upon phase transition. Linear stability analysis establishes instability thresholds of circles with no imposed twist, dependent on the ratio $q$ of the intrinsic curvature to the inverse radius of the original circle. Twisted circles are proven to be absolutely unstable but the linear analysis well predicts the dependence of the looping number of the emerging configurations on the imposed twist. Modeling stable configurations by relaxing numerically the overall elastic energy detects multiple stable and metastable states with different looping numbers. The bifurcation of untwisted circles turns out to be subcritical, so that nonplanar shapes with a lower energy exist at $q$ below the critical value. The looping number of stable shapes generally increases with $q$.

\end{abstract}


\end{frontmatter}


\section{Introduction} 

Studies of slender elastic rods or filaments possessing intrinsic curvature go back to Timoshenko plates \cite{Timoshenko} bending due to non-uniform thermal expansion. In Nature, the various forms emerge due to intrinsic curvature during plant growth \cite{GorielyGoldstein,GorielyTabor}. Intrinsic curvature arises as well when a prestressed and an unstressed rods are glued together.  The theory of such ``birods" concentrated on their buckling resulting in variegated helical shapes and hemihelices containing helicity-reversing perversions \cite{Liu15,Audoly17,Goriely17}. 

Instabilities of \emph{closed} twisted loops have a long history reviwed by Goriely \cite{Goriely06}, starting from the early work by Michell \cite{Michell}. The problem of closed loops with intrinsic curvature has been extensively studied in the relation to DNA configurations \cite{DNA,Tobias} where configurations of coiled and twisted closed loops are of particular interest for understanding the packing and reading genetic material. This prompted modeling and simulation of the various folding patterns emerging as a result of instabilities \cite{White}. 

In this paper, we consider configurations of another kind of elastic filaments with an emerging intrinsic curvature. Those are \emph{Janus filaments}, which can be fabricated by using two connected extruders with simultaneous melt spinning to generate a filament containing two different materials, say, a nematic and isotropic elastomers \cite{IonovJanus}. Extension or contraction of the active component made of a longitudinally polarized nematic elastomer due to a change of the nematic order parameter causes such a composite filament to acquire a wide variety of shapes, some of which have been recently investigated as key components of \emph{active textiles} \cite{azlp}. Alternatively, an elastic hydrogel \cite{Balazs} may serve as the active component. In this case, elongation or shortening is caused by local swelling or expelling the solvent. The two cases differ only by a change of the filaments cross-section, which grows in the same proportion as the length in the hydrogel but shrinks to preserve the volume in nematic elastomers. We will further concentrate on the nematically actuated filaments but only slight changes have to be made to adjust to the swelling case. Other ways of activation, \emph{e.g.} electrostriction may be employed. 

We will concentrate on spontaneous coiling and twisting of a single closed Janus loop. In the theoretical perspective, this case differs from those considered before in one important aspect: since the difference between the active and passive components is \emph{material} rather than being caused by a differential prestress, both components remain stressed in an equilibrium configuration, so that the overall elastic energy never vanishes.  After defining the problem in Sect.~\ref{S1}, we study in Sect.~\ref{S2} instabilities of a Janus circle. The linear analysis in Sect.~\ref{S21}, based on intrinsic equations of motion of curves, detects instability thresholds of untwisted circular filaments at different wavelengths quantized due to the periodicity. Twisted Janus filaments are proven in Sect.~\ref{S22} to be absolutely unstable, unlike twisted rods without intrinsic curvature that exhibit a sequence of bifurcations \cite{GorielyTabor}. 

Large-amplitude deformations, studied by relaxing numerically the overall elastic energy and described in Sect.~\ref{S3}, show that the bifurcation of nonplanar states detected by the linear theory is subcritical, so that non-planar configurations with different looping numbers coexist with the planar shape in the subcritical region. Among these configurations, those with a larger looping number become preferred as the intrinsic curvature increases but those with a lower looping number may persist as metastable states. Twisted filaments acquire a wide variety of shapes with a looping number fitting the imposed twist, at least at small to moderate ratios of the intrinsic curvature to the inverse radius of the original circle.    

\section{Energy of a slender Janus filament \label{S1}} 

We consider a Janus filament of a radius $r$ that consists of an active component occupying the sector $|\phi|<\pi/2$ of its circular cross-section and a passive component in the sector $|\phi|>\pi/2$. The active component of the filament is assumed to be a nematic elastomer that polarizes upon phase transition along the filament axis thereby causing elongation by a factor $\lambda=1+\epsilon$, while the passive component remains unchanged.   

When the nematic component elongates, the centreline of the filament develops an intrinsic curvature, which, when unforced and unconstrained, is directed along the normal to the dividing plane. The curvature radius is proportional to $\epsilon r$ \cite{azlp}, and therefore small extensions are sufficient to strongly bend a thin filament. This direction may, however, change in the presence of constraints, in particular, when the filament is clamped or has a form of a closed loop. We shall see that a spontaneous twist may arise under these conditions.  

The elastic energy $\mathcal{F}$ of an inextensible filament can be written in the slender body (Kirchhoff rod) approximation as the integral over the arc length $s$ of its neutral line:
\begin{align}
&\mathcal{F} =  \frac12 A E \int \left( \mathcal{F}^b + \mathcal{F}^t \right) d s, \notag\\
& \mathcal{F}^b  =  I(\kappa,\theta)   , \quad
 \mathcal{F}^t =  J\widehat{\tau}^2  ,
\label{Fe1}
\end{align}
where $A$ is the cross-sectional area, $E$ is the Young modulus, and $u$ is the axial strain of the centreline; $\kappa=1/R$ is the curvature of the centreline with the curvature radius $R$ and $I(\theta)$ is the flexural rigidity dependent on the angle $\theta$ between the normal to the dividing plane (to be called internal normal) and the curvature vector \textbf{n}, and $J=r^2/2$ is the torsional rigidity. The twist $\widehat{\tau}=\tau+d\theta/ds$ coincides with the geometric torsion $\tau$ in the Frenet frame only when the internal normal does not deviate from the normal vector.

The flexural rigidity is computed by adding the contributions of the active and passive sectors:
\begin{align}
&I_a =\frac{1}{\pi r^2}\int_0^r r \,d r 
\int_{-\pi/2}^{\pi/2} \left(\frac rR \cos (\phi-\theta) -\epsilon\right)^2 d \phi  = \frac 18  \left(\frac{r}{R}\right)^2 - \frac{4\epsilon}{3\pi}\frac{r}{R}\cos\theta +  \frac{\epsilon^2}{2} ,
  \label{Iin} \\
&I_p =\frac{1}{\pi r^2}\int_0^r \frac {r^3}{R^2} \,d r  
\int^{3\pi/2}_{\pi/2}\cos^2 (\phi-\theta) \,d \phi
= \frac 18  \left(\frac{r}{R}\right)^2, \label{Iout} \\
&I (\kappa,\theta) = I_a+I_p = \frac 14  \left(\kappa r\right)^2 
 -\frac {4\,\epsilon}{3 \pi} \kappa  r \cos\theta+\frac {\epsilon^2}{2} ,
 \label{Ipi}
\end{align}
The lowest energy state is clearly attained at $\theta=0$ when the normal to the dividing surface (internal normal) is directed along the normal vector $\mathbf{n}$ in the Frenet frame (see Appendix). The intrinsic curvature $\widehat{\kappa}$  is defined by the condition $dI/d\kappa=0$, yielding  
 \begin{equation}
\widehat{\kappa} r= \frac 83 \frac\epsilon\pi \label{khat}
\end{equation}
The energy of an inextensible  filament is expressed using Eqs.~\eqref {Ipi}, \eqref {khat} in Eq.~\eqref {Fe1} and omitting the constant term as 
 \begin{equation}
\mathcal{F} = \frac12 A Er^2 \int \left[ \frac14 \kappa(\kappa - 2\widehat{\kappa}
 \cos\theta ) + \frac{1}{2}\widehat{\tau}^2 \right]d s .
 \label{ekt}
\end{equation}
This formula does not coincide with the standard expression proportional to $(\kappa - \widehat{\kappa})^2$ \cite{AP}, which vanishes at $\kappa =\widehat{\kappa}$, whereas a Janus filament  remains under compression even when optimally bent. The above formula with modified numerical coefficients and appropriate values of  $\widehat{\kappa}$ is applicable also to filaments with different cross-sectional shapes and distributions of active and passive components. 

Equilibrium configurations minimize the total energy of the system, and can be attained following the pseudo-time evolution equations for the position of centerline of the filament $\mathbf{x}(s)$ and the direction of the internal normal:
 \begin{equation}
\frac{d \mathbf{x}(s)}{d t}= - \frac{\delta\mathcal{F}}{\delta \mathbf{x}(s)},
  \qquad \frac{d \theta(s)}{d t}= - \frac{\delta\mathcal{F}}{\delta \theta(s)}.
 \label{evol}
\end{equation}
These equations are discretized as 
 \begin{equation}
\frac{d \mathbf{x}_{i}}{d t}= - \frac{\partial}{\partial \mathbf{x}_{i}} 
 \sum_\mathrm{i}(\mathcal{F}^{s}_{i}+\mathcal{F}^{b}_{i} ),
 \qquad
\frac{d \theta_{i}}{d t}= - \frac{\partial}{\partial \theta_{i}}   
 \sum_\mathrm{i}(\mathcal{F}^{b}_{i}+\mathcal{F}^{t}_{i}) .
 \label{evold}
\end{equation}
The strain energy $\mathcal{F}^s = u_s^2$ is added here to suppress axial extension $u_s$ at $r\ll 1/\kappa$. While evolving the nodes on the centerline according to these equations, we track their positions $\mathbf{x}_i(s)$ to avoid self-intersections at points far removed along the arc length, and therefore knotted configurations are excluded in further computations.  

\section{Stability analysis\label{S2}}

\subsection{Energy increment due to virtual displacements\label{S20}}

The shape of a filament can be presented in the coordinate-independent form with the help of intrinsic equations of curvature and torsion in the Frenet frame (see Appendix). Using these equations to evolve the filament is not practical but they are advantageous for locating instabilities of an original shape by expressing with their help the change of the energy under virtual normal and binormal displacements $v,w$: 
 \begin{align}
\delta\mathcal{F} &= \frac14A E r^2 \int \left[ \left(  \kappa - \widehat{\kappa} \right)\delta\kappa  + \kappa \widehat{\kappa}\sin\theta \,  \delta\theta \right. + \left. 2(\tau \, \delta \tau  + \tau \, \delta \theta_s + \theta_s \delta \tau + \theta_s \delta \theta_s) \right] d s.
 \label{Ft}
\end{align}
where the subscripts denote derivatives.

The explicit form of  Eqs.~(\ref{Lkt}), (\ref{Ltt}), containing also the axial velocity $u$ that takes care of the incompressibility constraint, is, omitting terms of higher-order in small displacements that are irrelevant to further analysis,
 \begin{align}
\delta\kappa  &=  v_{ss} +\kappa^2 v +u \kappa_s + \kappa u_s  -\tau_s w -2 \tau w_s,  
\label{evolk} \\
 \delta\tau  & = \kappa w_s   +  \kappa^{-1}w_{sss}.
 \label{evolt} 
\end{align}
In the following, we investigate linear stability of a circle with the curvature $\kappa_0$,  which may be also twisted, to harmonic perturbations
 \begin{equation}
 v=\widetilde{v}\E^{\I k s}, \quad w=\widetilde{w}\E^{\I k s},\quad
u =\widetilde{u}\E^{\I k s}, \quad \delta\theta =\widetilde{\theta}\E^{\I k s},
\label{linu} 
\end{equation}
where the wavenumber $k=n \kappa_0$ should be an integer multiple of $\kappa_0$. The change of energy $\delta\mathcal{F}$ is quadratic in the perturbation amplitudes $\widetilde{v}, \widetilde{w}, \widetilde{u}, \widetilde{\theta}$, and the basic configuration is  unstable when $\delta\mathcal{F}<0$.  

\subsection{Stability of an untwisted filament\label{S21}}

We start with testing stability of an untwisted ($\tau_0=0$) circular filament with the curvature $\kappa_0$. For this end, we need to compute the lowest order (quadratic) resonant terms in Eq.~\eqref {Ft} to determine whether the energy decreases upon a virtual small displacement of the filament. In the lowest energy configuration $\theta_0=0$, so that the deviations are $\delta\tau=\tau, \, \delta\theta= \theta$, and can be presented in a harmonic form
 \begin{equation}
  \tau=\widetilde{\tau}\E^{\I k s} + \widetilde{\tau}^*\E^{-\I k s}, \quad
   \theta=\widetilde{\theta}\E^{\I k s} + \widetilde{\theta}^*\E^{-\I k s}, 
 \label{thint} \end{equation}
 where the asterisks marks the complex conjugates. The curvature is expressed as  
 \begin{equation}
 \kappa=\kappa_0+\widetilde{\kappa}\E^{\I k s}+\widetilde{\kappa}^*\E^{-\I k s} + \overline{\kappa},
 \label{lint} \end{equation}
where $ \overline{\kappa}$ denotes constant terms quadratic in displacements. The relations between the perturbation amplitudes follow from the linearized Eqs.~\eqref {evolk}, \eqref {evolt}:
\begin{equation}
\widetilde{\kappa}=- (k^2-\kappa_0^2) \widetilde{v}, \qquad
\widetilde{\tau}=- \frac{\I k} {\kappa_0} (k^2-\kappa_0^2) \widetilde{w}.
\label{kapv}  
\end{equation}
In the lowest order, the tangential displacement amplitude is expressed using Eq.~\eqref {Lgt} as $\widetilde{u} = -\I( \kappa_0 /k)\widetilde{v}$. Using this together with Eqs.~\eqref {linu}, \eqref {lint} in Eq.~\eqref {evolk} yields the second-order resonant term 
 \begin{equation}
\overline{\kappa}  =  2 (k^2/\kappa_0) (k^2 - \kappa_0^2 )|\widetilde{w}|^2. 
\label{evolk2}
\end{equation}

Substituting Eqs.~\eqref {kapv}, \eqref {evolk2} in Eq.~\eqref {Ft} yields the following second-order resonant terms in the integrand: 
 \begin{align}
 \delta\overline{\mathcal{F}} &= (\kappa_0 - \widehat{\kappa})\overline{\kappa}
 + 2|\widetilde{\kappa}|^2 
 +2 ( \kappa_0  \widehat{\kappa}+2k^{2}) |\widetilde{\theta}|^2   +4|\widetilde{\tau}|^2 + 2\I  k(\widetilde{\tau}^*\widetilde{\theta}-\widetilde{\tau}\widetilde{\theta}^*).  \label{Ft2}
\end{align}
Normal displacements affect only the second term of this expression, which reduces to $\kappa_0^4(n^2 - 1 )^2|v|^2>0$.  Hence, the filament is always stable to normal displacements.

Instability to binormal displacements may be caused either by the first term in Eq.~\eqref {Ft2}, which is negative at $\widehat{\kappa}>\kappa_0$, $n>1$, or by the last term dependent on the phase difference between the binormal displacement and rotation of the internal normal. The latter term can be presented using Eq.~\eqref {kapv} as $-4\kappa_0^3n^2(n^2-1)|\theta w|\cos \psi$, where $\psi$ is the difference between the phases of $\widetilde w$ and $\widetilde \theta$. Thus, the instability is most likely when the binormal displacement and rotation of the internal normal are \emph{in phase}. Setting $ \psi =0$, omitting the $v$-dependent term, and denoting $ q=\widehat{\kappa}/\kappa_0, \, \widetilde\vartheta = \widetilde{\theta}/\kappa_0 = \zeta \widetilde w$, Eq.~\eqref {Ft2} is rewritten as 
\begin{align}
  \delta\overline{\mathcal{F}}&= 2 \kappa_0^4 \left[n^2(n^2-1)(2n^2-q-1) \right.  + \left. (q+2n^2) \zeta^2 - 2 n^2(n^2-1) \zeta \right] |\widetilde w|^2. 
  \label{Ft3}
\end{align}
This expression is at minimum with respect to the amplitude ratio $\zeta$ at $\zeta_c=n^2(n^2-1)/(q+2n^2)$. Using this in Eq.~\eqref {Ft3}, it is easy to see that zeroes of $\delta\overline{\mathcal{F}}$ satisfy the quadratic equation $q(1-q)=n^2(3n^2-1)$, which yields the instability threshold   
\begin{equation}
 q_c = \frac 12 \left(\sqrt{1 - 4 n^2 + 12 n^4} -1\right). 
  \label{Ftq}
\end{equation}
The numerical values are $\zeta_c \approx 0.848,  \, q_c \approx6.152$ at $n=2$, $\zeta_c \approx 2.195, \, q_c \approx 14.8$ at $n=3$, $\zeta_c \approx 4.073, \, q_c \approx 26.9$ at $n=4$.

\subsection{Instability of a twisted filament\label{S22}}

Next we consider the original configuration with an imposed twist $\widehat{\tau}_0=m \kappa_0$ where $m$ must be an integer due to the periodicity. In a twisted filament, the internal normal deviates from the radial direction, so that $\theta_0=\int \widehat{\tau}_0 ds = m\varphi$ depends linearly on the angular coordinate $\varphi =\kappa_0 s$, while the torsion of the Frenet frame $\tau_0=0$. Clearly, this greatly increases the energy of the twisted configuration through the term containing $\cos\theta$ in Eq.~\eqref {ekt}. The twist of a circular filament cannot be eliminated by topological reasons but the energy can be relaxed by a \emph{finite} $m$-periodic resonant perturbation $\delta\theta = a \sin m\varphi $. The increment of the integral in Eq.~\eqref{ekt} is then
\begin{equation}
\widehat{\kappa} \int_0^{2\pi} \cos(m\varphi +a \sin m\varphi) \D\varphi =
- 2\pi \widehat{\kappa} J_1 (a),
  \label{Ftt}
\end{equation}
where $J_1 (a)$ is a Bessel function. The energy decrement is at maximum at $a=a_0 \approx 1.84$, independently of $m$. It is stability to small perturbations of the lowest energy state 
\begin{equation}
\widehat\theta= m \varphi +a_0 \sin m\varphi 
  \label{thhat}
\end{equation}
that has to be further investigated. 

The energy increment Eq.~\eqref {Ft} depends now linearly on complex displacement amplitudes, and it can be manipulated in a desired way by suitably adjusting the phases of perturbations; thus, we expect a twisted filament to be always unstable. 
For a formal proof, it suffices to compute lowest order perturbations of the $\theta$-dependent terms in Eq.~\eqref {ekt} using there the $\widetilde\kappa,\,\widetilde\tau$ given by the linearized relations \eqref {kapv}:
\begin{align}
\delta{\mathcal{F}}&=\kappa_0 \int_0^{2\pi}\left(\frac q2 \widetilde\kappa
\cos \widehat \theta +\widetilde\tau  \widehat \theta_\varphi \right)\D\varphi  =  \kappa_0^3(n^2-1)\left(- q \mathcal{I}_v |\widetilde v| 
   +2n \mathcal{I}_w |\widetilde w| \right),
  \label{dtau1}\\
  \mathcal{I}_v&=\int_0^{2\pi} \cos (n \varphi+\psi_v) \cos (m \varphi +a_0 \sin m\varphi)\D\varphi,  \notag \\
 \mathcal{I}_w&=a_0 \int_0^{2\pi} \sin (n \varphi+\psi_w) \cos m\varphi\,\D\varphi 
 =\pi a_0 \sin \psi_w \delta_{mn}, \notag   
 \end{align}
where $\psi_v, \psi_w$ are phases of  the displacements $\widetilde v, \widetilde w$ and $\delta_{mn}$ is the Kronecker delta. The last formula indicates instability to resonant binormal displacements with $\sin \psi_w <0$, which is at the maximum at $\psi_w=-\pi/2$. The integral $ \mathcal{I}_v$ is evaluated as $A_{mn} \cos \psi_v $ with $A_{mn}=2\pi a^{-1} J_1 (a) \approx 1.986$ at $m=n$, $A_{mn} =2 \pi [J_1( a) -2 a^{-1} J_2( a]\approx 1.499$ at $n=2m$. It remains positive at $n=bm$ with an integer $b$ but decreases as $b$ grows, and vanishes otherwise at $m>1$. At $m=1$, $A_{1n}$ is positive at any $n$ and decreases as $n$ grows. These results indicate preferential instability at $\psi_v=\pi$, i.e. in antiphase to $\theta$. Thus, a twisted filament is always unstable to both normal and binormal displacements with suitable phases. The values of $A_{mn}$ suggest preferential development of forms with a fitting looping number, as is further confirmed by the simulations in Sect.~\ref{S32}.

\section{Large-amplitude deformations\label{S3}}

\subsection{Deformations of untwisted loops \label{S31}}

\begin{figure}[t]
\centering
\begin{tabular}{cc}
 (a) & (b)\\
 \includegraphics[width=.48\textwidth]{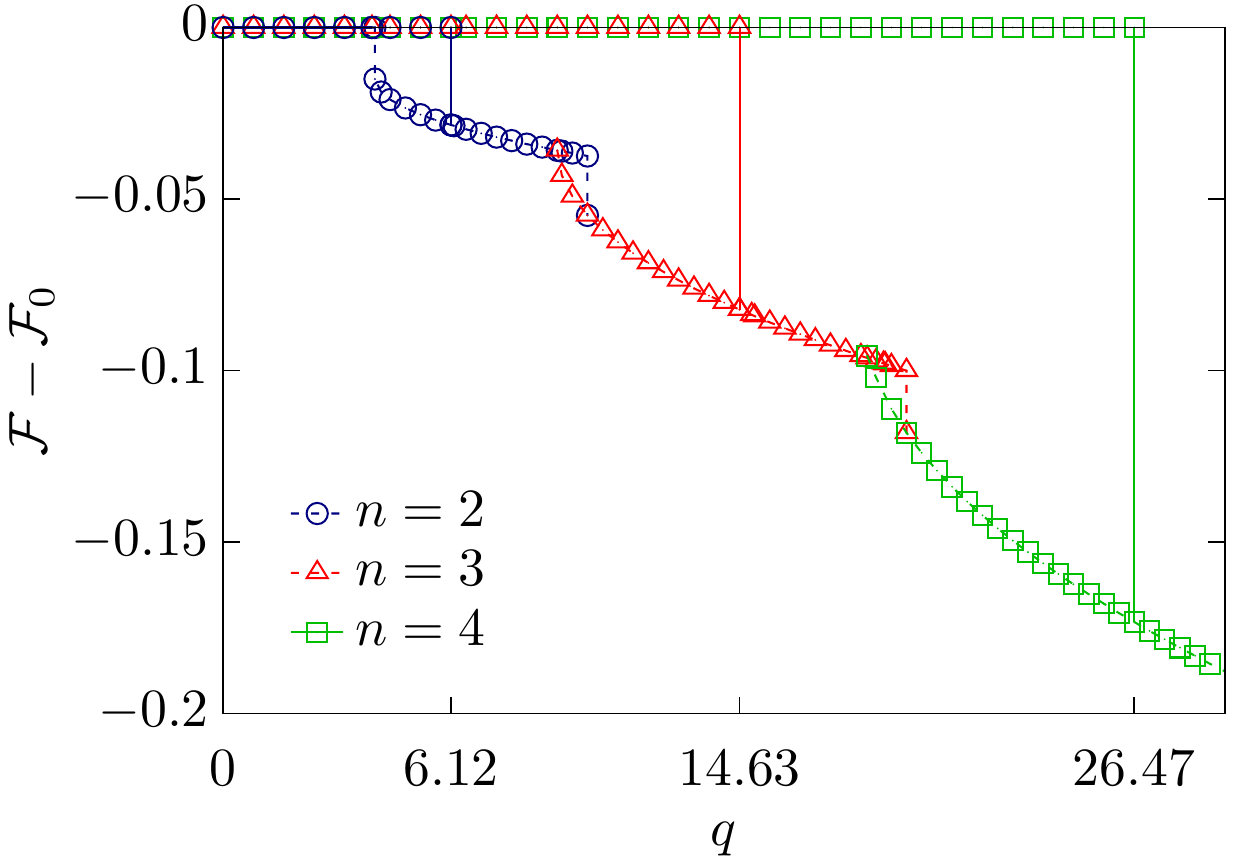}
&\includegraphics[width=.49\textwidth]{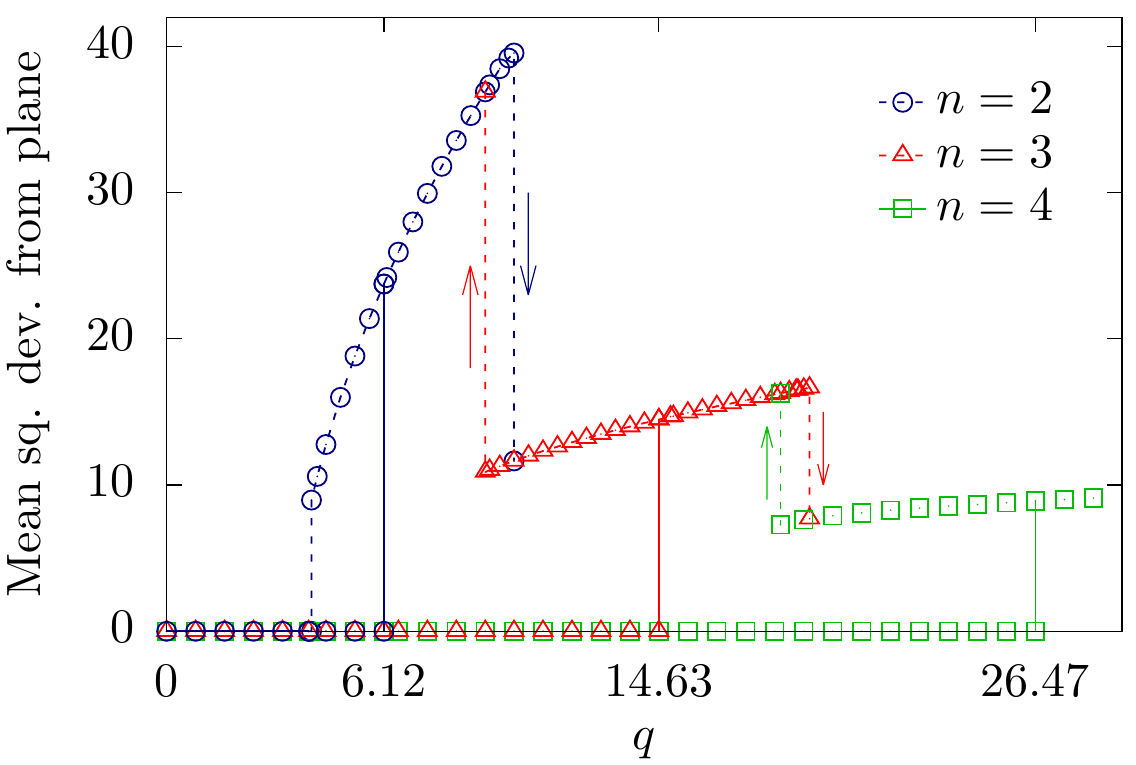}
\end{tabular}
\caption{The dependence of quantitative characteristics of non-planar configurations with different looping numbers $n$ on the parameter $q$. (a): The energy decrements relative to the planar loop. (b): The mean square deviation from the original plane.}  
\label{Energy} \end{figure}

Computation of large-amplitude structures emerging due to instability of untwisted Janus loops shows that the bifurcations described in Sect.~\ref{S21} are \emph{subcritical}, and three-dimensional configurations with lower energy exist already at $q<q_c$. The dependences of the energy decrements relative to the planar circle and of the mean square deviation from the original plane on $q$ are shown in Fig.~\ref{Energy}. 
\begin{figure}[t]
\centering
 \includegraphics[width=.31\textwidth]{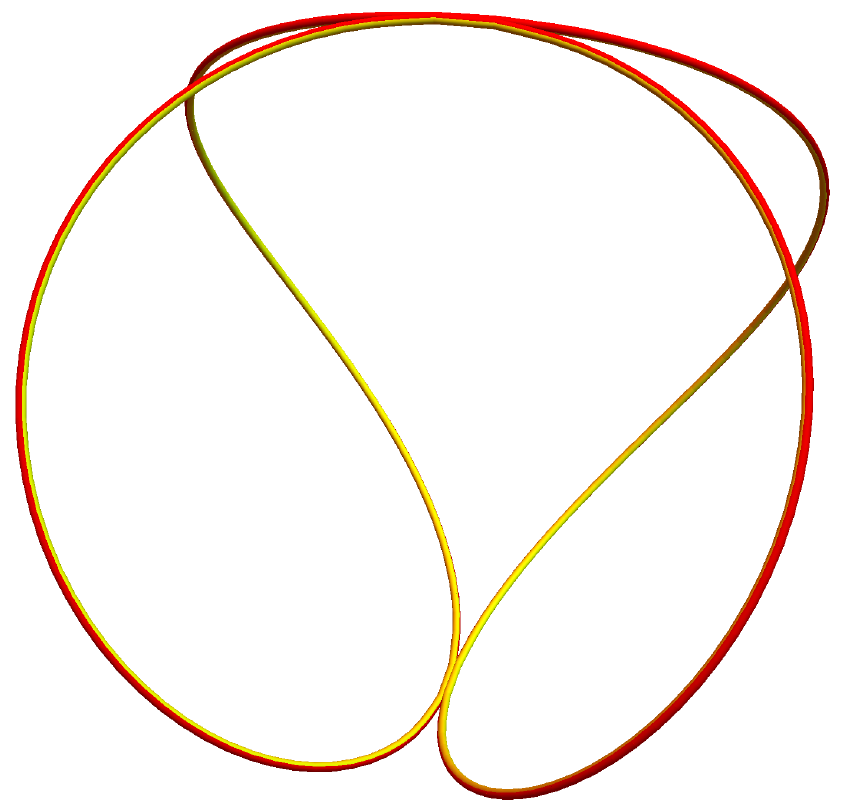}
\includegraphics[width=.31\textwidth]{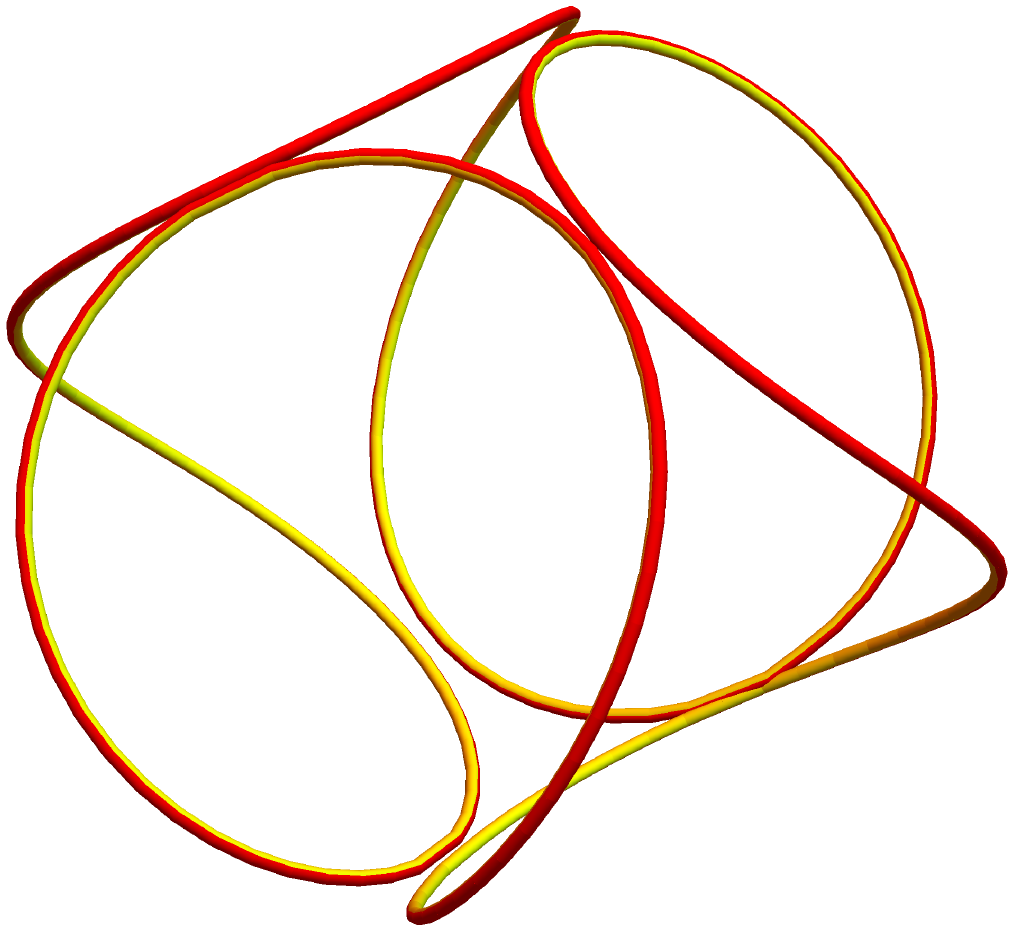}
\includegraphics[width=.31\textwidth]{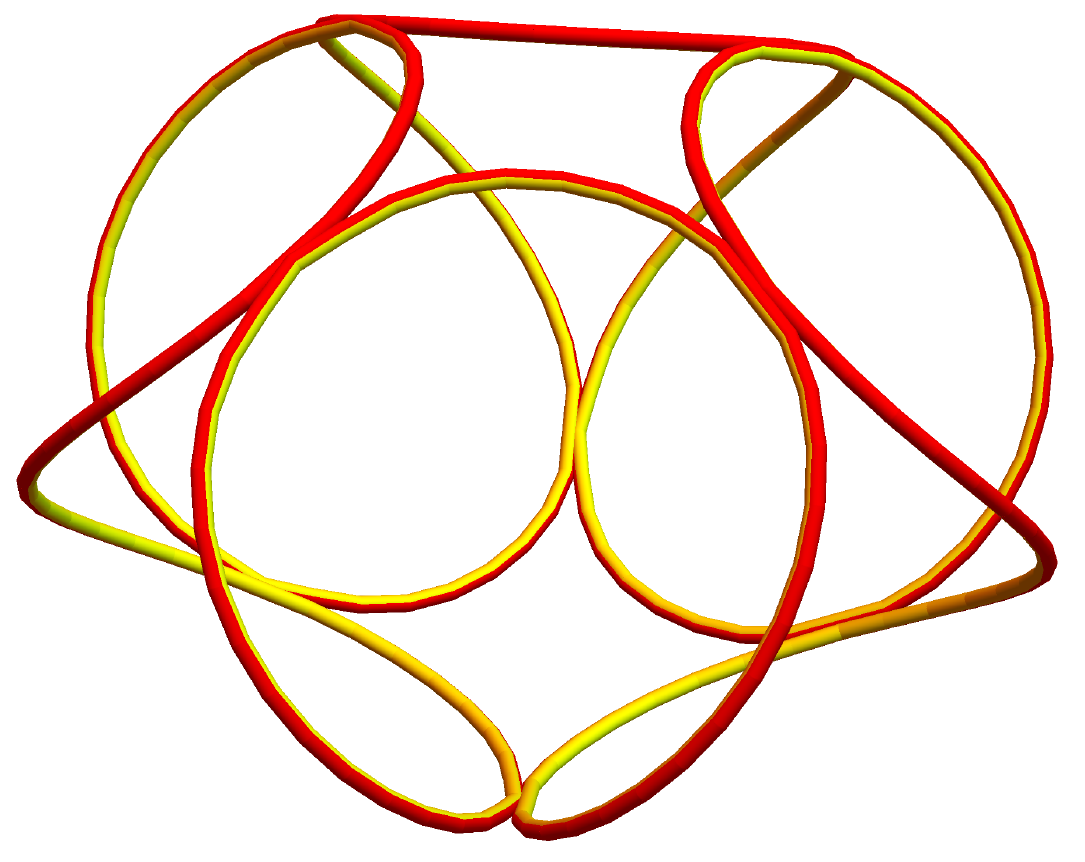}
\caption{Configurations with the looping numbers $n=2$  at $q=9$ (left), $n=3$  at $q=16$ (middle), and $n=4$  at $q=27$ (right). The filament radius and length are the same in all cases. The active and passive components are colored in red (dark) and yellow (light) respectively. }  
\label{Shape} 
\end{figure}

As $q$ grows, configurations with an increasing looping number $n$ appear; a sample of shapes is shown in Fig.~\ref{Shape}. The energy plots in Fig.~\ref{Energy}a show that the looping number of the lowest energy configuration increases with $q$; higher-energy configurations remain, however, metastable within a certain interval and can be attained from the original planar configuration when perturbing it at a suitable wavelength. In particular, the planar state is recovered at $q<q_c(n)$ when perturbations with the same $n$ are sufficiently small. According to Fig.~\ref{Energy}b, the deviations from the original plane tends to decrease with the increasing looping number.

Since our computation procedure is based on minimization of energy, we cannot compute unstable configurations serving as basin boundaries between the various absolutely stable and metastable states. The evolution to the lowest-energy form commonly terminates when the filament reaches a self-contact of  points far removed along the curve. We checked whether a slight asymmetric perturbation might help to avoid self-contact to further decrease the energy, but the same configurations were recovered.

\subsection{Deformations of twisted loops \label{S32}}

\begin{figure}[t]
\centering
\begin{tabular}{cc}
 (a) & (b)\\
 \includegraphics[width=.48\textwidth]{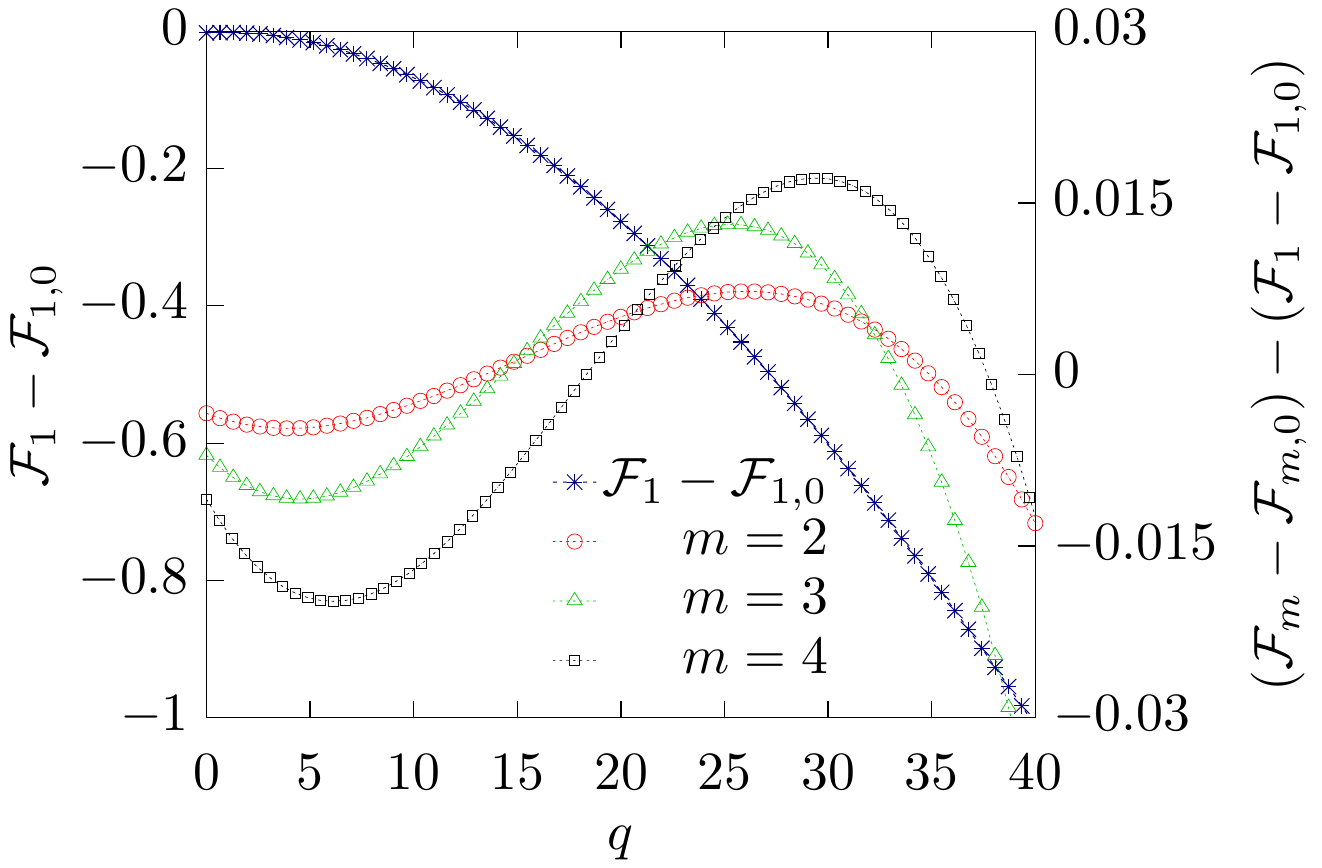}
&\includegraphics[width=.49\textwidth]{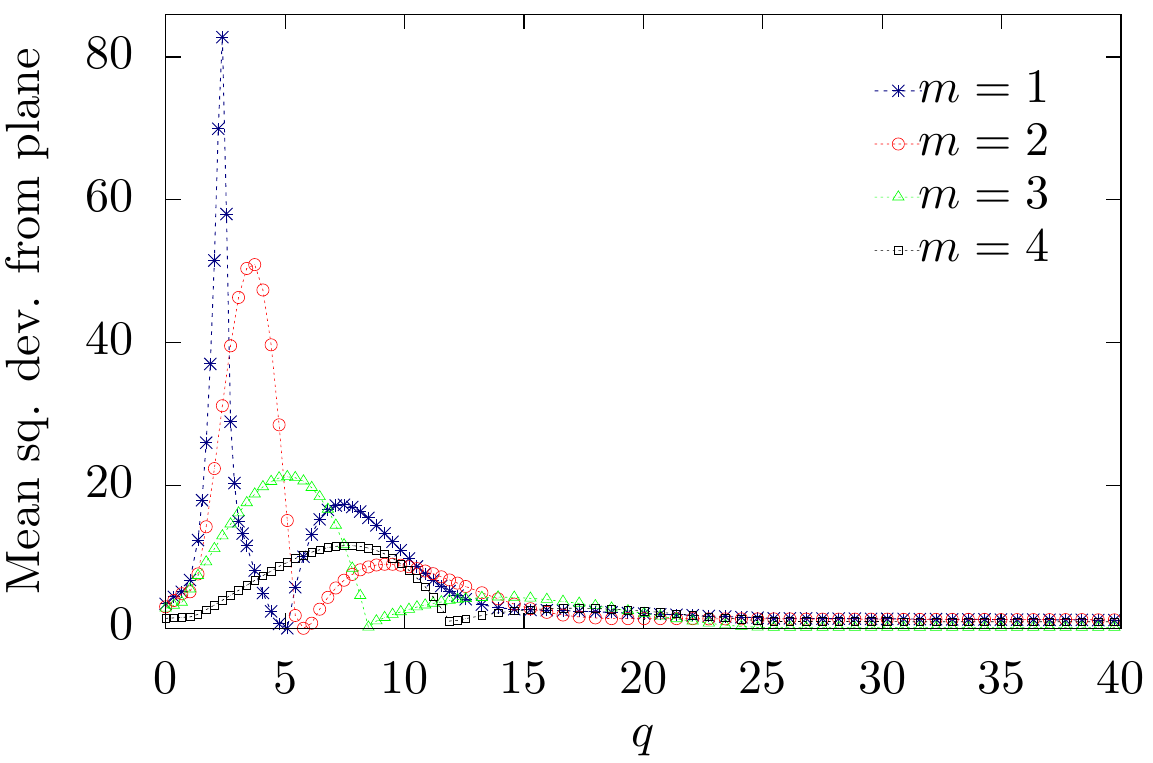}
\end{tabular}
\caption{Dependences of the energy gain comparative to initial state (a) and the mean square deviation from the original plane (b) on the parameter $q$ for different twist numbers $m$.}  
\label{EnergyTwisted} \end{figure}

In agreement with the stability analysis in Sect.~\ref{S22}, non-planar configurations with different looping numbers, dependent on the imposed twist, appear already at small $q$. We have not tested specific effects of perturbations on different wavelengths but allowed the shape to develop in a natural way choosing an optimal looping number. The prediction of the above analysis is that a perturbation on the wavelength with the largest value of the coefficient $A_{mn}$ is preferred at small amplitudes. This is, indeed, what has been observed in the simulations, as it seen in Fig.~\ref{EnergyTwisted}a. At $m=1$, the emerging form at $q<5$ had the lowest non-trivial looping number $n=2$, and at higher $m$ the looping number $n=m$ was selected at moderate values of $q$. Of course, the analytical predictions cannot be extended to large $q$ when large deviations from the planar shape promptly develop.      

The plots of the energy gain \emph{vs.} $q$ for the optimal configurations at different $m$ are very close one to the other, and for the sake of clarity we show in Fig.~\ref{EnergyTwisted}a only the curve  $\mathcal{F}_1-\mathcal{F}_{1,0}$ for $m=1, \, n=2$, while other plots show the difference between $\mathcal{F}_m-\mathcal{F}_{m,0}$ for $n=m$ and $\mathcal{F}_1-\mathcal{F}_{1,0}$.  

\begin{figure}[t]
\centering
\begin{tabular}{cccc}
 (a) & (b) & (c) & (d)\\
\includegraphics[width=.22\textwidth]{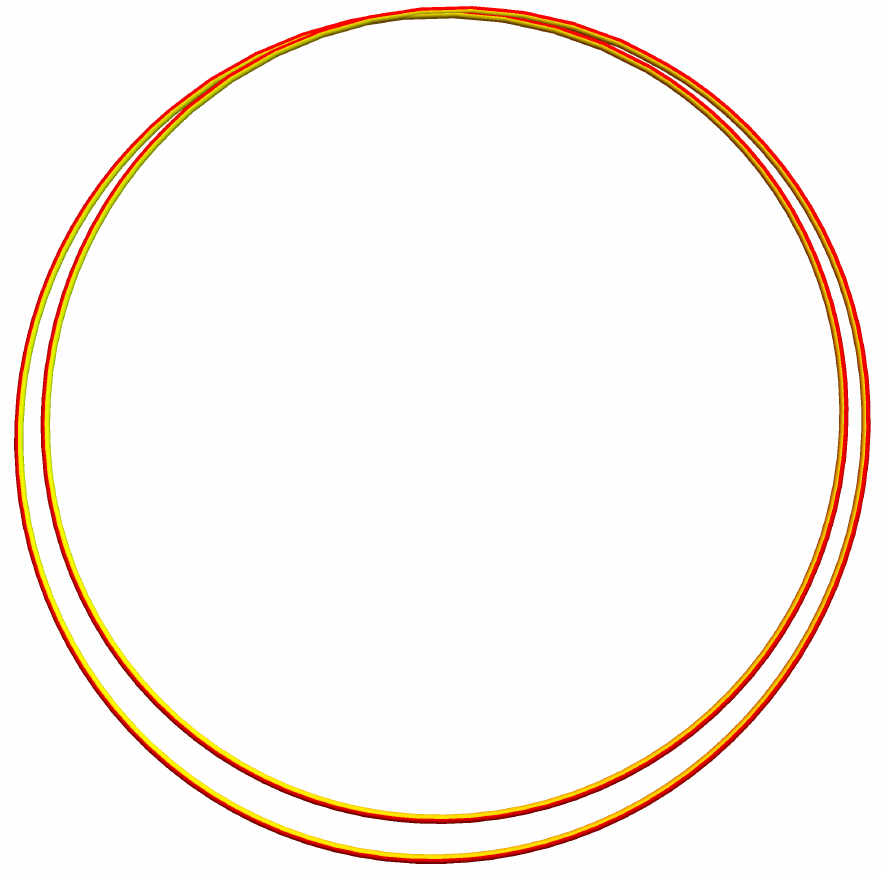}
&\includegraphics[width=.22\textwidth]{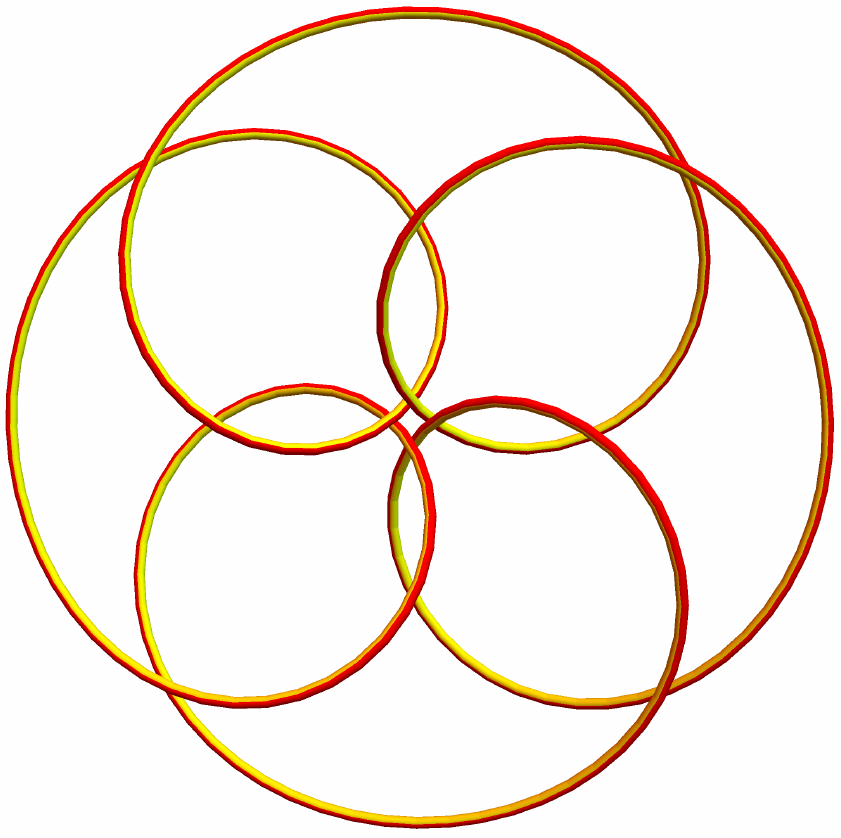}
&\includegraphics[width=.22\textwidth]{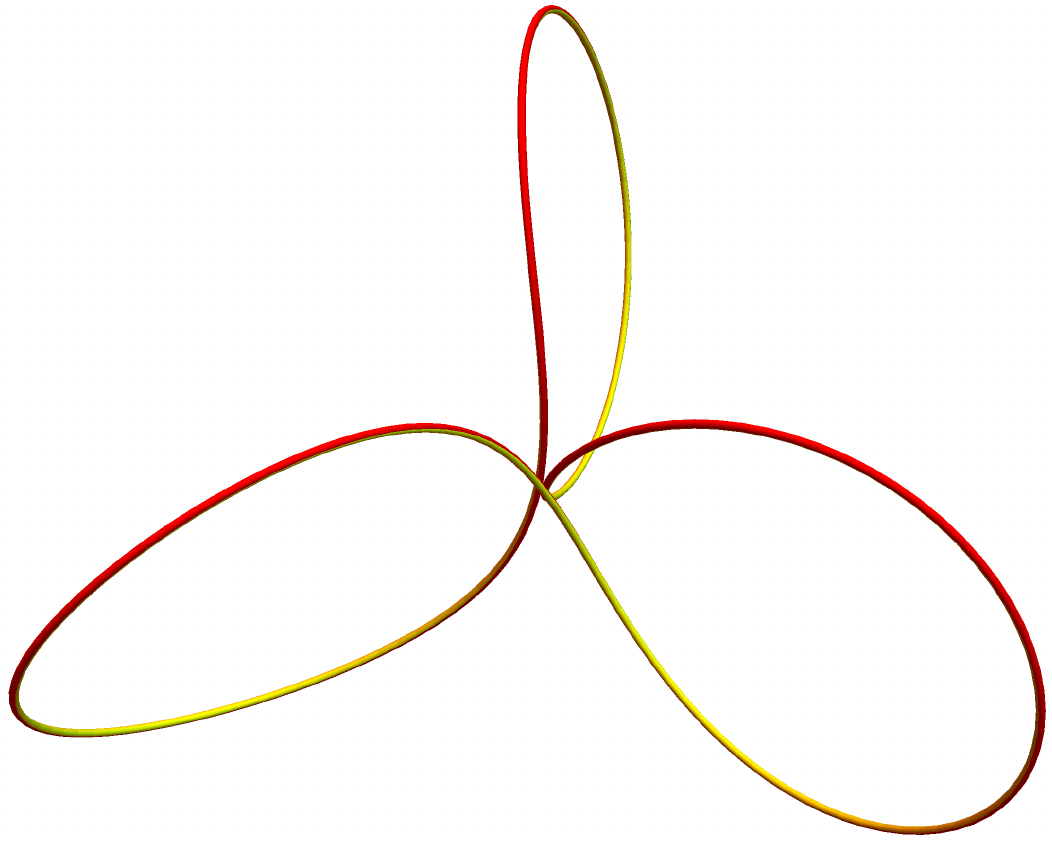}
&\includegraphics[width=.22\textwidth]{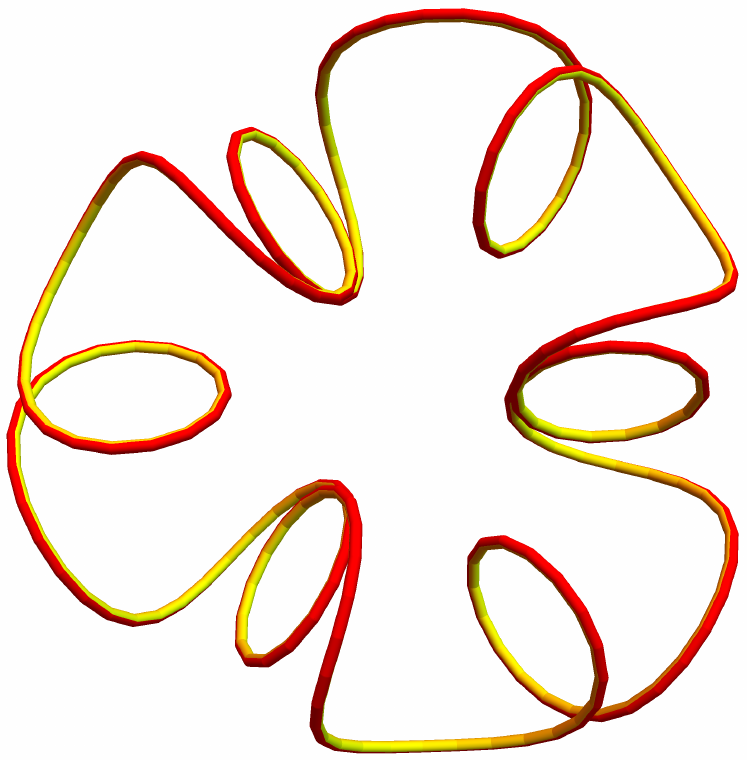}
\end{tabular}
\caption{Reshaping twisted ring filaments leading to equilibrium planar solutions at (a) $m=1$, $q=5$ (b) $m=4$, $q=12$ and to non-planar conformations with (c) single ($m=3$, $q=5$) and (d) multiple ($m=3$, $q=20$) self-contacts.}  
\label{ShapeTwisted} 
\end{figure}

The shapes, some of which are shown in Fig.~\ref{ShapeTwisted}, are extremely variegated, and the mean square deviation from the original plane strongly but non-monotonously depends on $q$, as seen in Fig.~\ref{EnergyTwisted}b. Each curve plotted in this Figure has two maxima with the height decreasing with both $m$ and $q$, separated by a dip where the mean square deviation drops to zero. We observe at this point a multiple coverage of the circle as, for example, a double coverage at $m=1$,  $q=5$ with one point of self-contact (Fig.~\ref{ShapeTwisted}a) and a more elaborate planar pattern at $m=4$, $q=12$ (Fig.~\ref{ShapeTwisted}b). Although planar shapes apparently exist at a certain $q$ at any $m$, generic shapes are non-planar and become more complicated with growing $q$. This is illustrated by a contrast between a trefoil with a self-contact in the centre at $m=3, \,q=5$ in Fig.~\ref{ShapeTwisted}c and a shape with multiple self-contacts and helicity-reversing perversions at $m=3, \,q=20$ in Fig.~\ref{ShapeTwisted}d. 

%


\section{Conclusion}

Actuation of Janus filaments can create in a controlled way a large variety of coiled and twisted shapes. We have investigated them here both analytically with the help of the intrinsic representation of deformation of curves, and numerically through relaxation toward energy minima, obtaining consistent results in the applicability limits of analytical theory. The specific property of Janus filaments, besides creating an intrinsic curvature that leads to variegated configurations of closed loops, is retaining compression even when optimally bent. Their configurations can be dynamically regulated by external controls affecting the intrinsic curvature through changes of the nematic order parameter. These properties may facilitate, in particular, reshaping in multiple ways active textiles \cite{azlp} framed or   embroidered by Janus filaments.   

\emph{Acknowledgement.} This research is supported by Israel Science Foundation  (grant 669/14).

\appendix\section{Derivation of intrinsic equations}

Consider a curve ${\cal C}$ defined in a parametric form ${\bf x}(s)$, where $s$ is the arc length. The unit tangent, normal and binormal vectors $\bf{l}, \bf{n}, \bf{b}$ forming the Frenet trihedron at any point of the curve are related by the Frenet--Serret equations
\begin{equation}
{\bf x}_s= {\bf l},  \;\;\; {\bf l}_s= \kappa {\bf n},  \;\;\;
{\bf n}_s= - \kappa {\bf l} + \tau {\bf b},  \;\;\; {\bf b}_s= -\tau {\bf n}.
\label{LFS}  \end{equation}
Let the curve be displaced by increments $u,v,w$ along, respectively, $\bf{l}, \bf{n}, \bf{b}$, so that  
\begin{equation}
\delta \mathbf{x} = u{ \bf l}+v { \bf n}+w{ \bf b}.
\label{Lvel}  \end{equation}
The tangential velocity is a gauge variable that expresses a  reparametrization of a displaced curve. For a nonextensible filament, differentiating the first Frenet--Serret equation with respect to time and replacing the derivatives with respect to $s$ with the help of other equations, yields the increment of {\bf l}: 
\begin{equation}
\delta{\bf l} =  \delta{\bf x}_s = (u_s  - \kappa v) {\bf l}  +   (v_s + \kappa u - \tau w) {\bf n} + (w_s + \tau v) {\bf b}.  
\label{Llt}  \end{equation}
Since {\bf l} is a unit vector, the projection $\delta\mathbf{l \cdot l}$ should vanish. This yields the relation between $u$ and $v$ in the isometric gauge
\begin{equation}
u_s = \kappa v. 
\label{Lgt}  \end{equation}
Equation (\ref{Llt}) thus reduces to
\begin{equation}
\delta{\bf l} = V {\bf n} + W{\bf b}; \;\;\; 
V =v_s + \kappa u - \tau w, \;\;\; W = w_s + \tau v.  
\label{Llt1}  \end{equation}
In the same manner, we obtain from the second Frenet--Serret equation with the help of Eqs.~(\ref{Lgt}) and (\ref{Llt1}) he increment of {\bf n}: 
{\samepage \begin{eqnarray}
\kappa\, \delta{\bf n} &= & - \delta \kappa\,{\bf n}
+ \partial_s (V {\bf n} + W{\bf b})  = - V {\bf l}+  (W_s +  \tau V) {\bf b} \nonumber \\
&+&  (-\delta \kappa  -\kappa u_s + \kappa^2 v +V_s - \tau W) {\bf n}.  
\label{Lnt}  \end{eqnarray}}
Again, requiring the projection $\delta\mathbf{n \cdot n}$ to vanish and using Eq.~(\ref{Lgt}) gives the increment of the curvature:
\begin{equation}
\delta\kappa  = V_s - \tau W. 
\label{Lkt}  \end{equation}
The evolution equation of {\bf n} reduces therefore to
\begin{equation}
{\bf n}_t = - V {\bf l} + U{\bf b}; \qquad
U = \kappa^{-1}( W_s +  \tau V).  
\label{Lnt1}  \end{equation}
Finally, the increment of the torsion $\tau$ is obtained from the third Frenet--Serret equation after using Eq.~(\ref{Lgt}) and requiring the projection $\delta{\bf b \cdot b}$ to vanish:
\begin{equation}
\delta\tau  =  \kappa W + U_s. 
\label{Ltt}  \end{equation}
%


\end{document}